\documentclass[traditabstract]{aa} 
\usepackage{graphicx}
\usepackage{txfonts}
\usepackage{natbib}
\bibpunct{(}{)}{;}{a}{}{,}

\begin{document}

   \title{Detection capability of ground-based meter-sized telescopes for shallow exoplanet transits
\thanks{Based on data obtained with the STELLA robotic telescopes in Tenerife, an AIP facility jointly operated by AIP and IAC.}}

   \titlerunning{Shallow exoplanet transits}

   \author{M.~Mallonn\inst{1}, K. Poppenhaeger\inst{1,2}, T. Granzer\inst{1}, M. Weber\inst{1}, K. G. Strassmeier\inst{1,2}}
   \authorrunning{M. Mallonn et al.}

\institute{
\inst{1}Leibniz-Institut f\"{u}r Astrophysik Potsdam, An der Sternwarte 16, D-14482 Potsdam, Germany \\
\inst{2}Institut f\"{u}r Physik und Astronomie, Universit\"{a}t Potsdam, Karl-Liebknecht-Straße 24/25, D-14476 Potsdam, Germany \\
\email{mmallonn@aip.de}
}

   \date{Received --; accepted --}

  \abstract{Meter-sized ground-based telescopes are frequently used today for the follow-up of extrasolar planet candidates. While the transit signal of a Jupiter-sized object can typically be detected to a high level of confidence with small telescope apertures as well, the shallow transit dips of planets with the size of Neptune and smaller are more challenging to reveal. We employ new observational data to  illustrate the photometric follow-up capabilities of meter-sized telescopes for shallow exoplanet transits. We describe in detail the capability of distinguishing the photometric signal of an exoplanet transit from an underlying trend in the light curve. The transit depths of the six targets we observed, Kepler-94b, Kepler-63b, K2-100b, K2-138b, K2-138c, and K2-138e, range from 3.9~ppt down to 0.3~ppt. For five targets of this sample, we provide the first ground-based photometric follow-up. The timing of three targets is precisely known from previous observations, and the timing of the other three targets is uncertain and we aim to constrain it. We detect or rule out the transit features significantly in single observations for the targets that show transits of 1.3~ppt or deeper. The shallower transit depths of two targets of 0.6 and 0.8~ppt were detected tentatively in single light curves, and were detected significantly by repeated observations. Only for the target of the shallowest transit depth of 0.3~ppt were we unable to draw a significant conclusion despite combining five individual light curves. An injection-recovery test on our real data shows that we detect transits of 1.3~ppt depth significantly in single light curves if the transit is fully covered, including out-of-transit data toward both sides, in some cases down to 0.7~ppt depth. For Kepler-94b, Kepler-63b, and K2-100b, we were able to verify the ephemeris. In the case of K2-138c with a 0.6~ppt deep transit, we were able to refine it, and in the case of K2-138e, we ruled out the transit in the time interval of more than $\pm\,1.5\,\sigma$ of its current literature ephemeris.  }


   \maketitle
%

\section{Introduction}

Space-based photometry of the Kepler satellite \citep{Borucki2010} has revolutionized the discovery of extrasolar planets. Current and future photometry satellites such as the Transiting Exoplanet Survey Satellite \citep[TESS,][]{Ricker2015} and the PLAnetary Transits and Oscillations of stars \citep[PLATO,][]{Rauer2016} will be the main instruments for exoplanet detection during their lifetime. These space missions are expected to find many sub-Neptune-sized and even Earth-sized exoplanets orbiting nearby bright host stars that are suitable for further characterization. Thus, after two decades of detailed investigation of large extrasolar gas giants, we will have the possibility to push toward detailed studies of smaller, observationally more challenging planets. In the detection and subsequent follow-up, ground-based telescopes play a significant role, and small-sized telescopes contribute as well with transit photometry to rule out false-positive scenarios and to refine the orbital ephemeris \citep[e.g.,][]{Kostov2019,Luque2019,Shporer2020}. The need for ground-based follow-up of TESS discoveries is explicit \citep{Dragomir2020,Kokori2021} in preparation of atmospheric characterization with the James Webb Space Telescope and the Atmospheric Remote-sensing Infrared Exoplanet Large-survey \citep[ARIEL,][]{Tinetti2018}. 

Ground-based, small-aperture telescopes are successfully employed in the follow-up of TESS planet candidates \citep[e.g.,][]{Kostov2019,Luque2019,Shporer2020,Cloutier2020a,Kemmer2020,Nowak2020}, but these studies typically rely on an ensemble of follow-up photometry and no emphasis is given to the detection significance of shallow transit features in an individual light curve. To estimate the transit depth that a typical ground-based observation is capable of detecting, the signal-to-noise ratio of a photometric transit observation can be approximated by the ratio of the transit depth over the photometric uncertainty of the data points and the square root of the number of data points within the transit duration \citep{Pont2006}.

In the past decade, one-meter telescopes routinely achieved 1-2~mmag precision in the point-to-point variation with a cadence of below 2 minutes \citep{Carter2011,Narita2013,Seeliger2015,Mallonn2016,Lendl2017,Delrez2018}. Thus, such simple estimation promises $3\,\sigma$ detections of transit depths down to 500~parts-per-million (ppm) for 1~mmag photometry and a 3-hour-long transit. However, hot-Jupiter transit observations rarely yield a per-light-curve $1\,\sigma$ precision in the transit depth of better than 300~ppm when it is obtained from the ground with small aperture telescopes \citep[e.g.,][]{Mallonn2015,Dragomir2015,Alexoudi2018}. Thus, 1~parts-per-thousand (ppt) is typically considered as a rough limit for a significant transit detection by individual light curves \citep{Lendl2017}. 

We wish to describe the photometric capabilities of a meter-sized telescope in the detection of transit signals of sub-Neptune-sized exoplanets. Earlier work on this topic has been presented by \cite{Lendl2017} and \cite{Smith2020} for small-sized telescopes, and by \cite{Stefansson2018} for a medium-sized telescope assisted by an engineered diffuser. Our contribution is the analysis of a large set of 19 new transit light curves by which we aim to obtain a representative conclusion. The majority of this new data is of high photometric quality, typical of what can be obtained today in routine operation. A few light curves present data of moderate quality to illustrate the drop in detection capabilities. We start the investigation with three targets from 0.8 to 3.9~ppt transit depth for which the timing of the transit signal is precisely known. Then, we move to objects with more uncertain timing predictions with a depth ranging from 0.3 to 1.3~ppt. For most objects, repeated observations were obtained, and we investigate the potential of single light curves as well as multiple light curves in a joint analysis.
In a last step, we further test the capabilities of our real data by an injection-recovery analysis of transit signals of very different properties.

In Section~\ref{sec_tar} we briefly introduce the chosen planetary targets, whose observations and data reduction are described in Section~\ref{sec_obs}. The results of the light-curve analysis is presented in Section~\ref{sec_ana}. They are discussed by a comparison to the results of an injection-recovery test in Section~\ref{sec_dis}. A summary is provided in Section~\ref{sec_concl}.

\begin{table*}
\caption{Physical and observational properties of the targets. The uncertainty of the predicted transit time $\Delta\,T_c$ is estimated for the first observation per target within this work.}
\label{tab_targets}
\begin{center}
\begin{tabular}{llccccccc}
\hline
\hline
\noalign{\smallskip}
Object & V  & Radius & Mass & Period & Depth & Duration & $\Delta\,T_c$ & Ref \\
& [mag] & [Earthradii] & [Earthmasses] & [days] & [ppt] & [hours] & [minutes] &  \\
\hline
\noalign{\smallskip}
Kepler-94b & 12.9 & 3.5 & 10.8 & 2.51 & 1.8 & 1.2 & 0.3 & 1,2 \\
Kepler-63b & 12.0 & 6.1 & <120 & 9.43 & 3.9 & 2.9 & 0.2 & 3,4 \\
K2-100b & 10.5 & 3.9 & 22 & 1.67 & 0.8 & 1.6 & 0.6 & 5 \\
K2-138b & 12.2 & 1.5 & 3.1 & 2.35 & 0.3 & 2.0 & 133 & 6 \\
K2-138c & 12.2 & 2.3 & 6.3 & 3.56 & 0.6 & 2.4 & 52 & 6 \\
K2-138e & 12.2 & 3.4 & 13.0 & 8.26 & 1.3 & 3.0 & 41 & 6\\
\hline  
\end{tabular}
\tablefoot{References. 1: \cite{Marcy2014}; 2: \cite{Holczer2016}; 3: \cite{SanchisOjeda2013}; 4: \cite{Gajdos2019}: 5: \cite{Barragan2019}; 6: \cite{Lopez2019}}
\end{center}
\end{table*}

\section{Targets}
\label{sec_tar}
Among all known transiting exoplanets, we selected targets in a range of transit depths around 1~ppt because this is indicated as a rough limit in the literature of the detection capabilities of small to medium-sized ground-based telescopes \citep{Lendl2017,Stefansson2018,Smith2020}. In an attempt to push the limits, we included targets with a transit depth of 0.6 and even 0.3~ppt. The transit duration of our targets is rather short, with 2~hours on average, to allow for more successful transit observations. Specifically, multiple light curves per target allow us to test whether results on individual targets can be reproduced, and to allow for the coaddition of multiple data sets per target to increase the sensitivity. The range of host star brightness of our targets from 10.5 to 12.9 V-band magnitude is typical for exoplanets detected from the ground. It also covers about 50\,\% of the TESS objects of interest (TOI), $\sim$\,20\% of which are brighter in V and $\sim$\,30\% are fainter \citep{Guerrero2021}. The parameters of all transit signatures searched for in this work have been determined precisely by observations with the Kepler photometric satellite, either in its prime or in its extended mission. For three of our targets, the timing of the transit events is also precisely known, which means that they provide a robust reference to which we can compare our results. For the other three targets, the timing is uncertain, and we attempt to refine it. Table~\ref{tab_targets} provides a summary of physical and observational properties of the six targets. A more detailed introduction of the individual exoplanets is given below. We note that for five of our six targets, this work constitutes the first ground-based follow-up observations.

\subsection{Kepler-94b}
The exoplanet Kepler-94b was detected by \cite{Marcy2014} in long-cadence data of the Kepler spacecraft. It was revealed by V-shaped transits with a depth of about 1.5~ppt, a duration of 1.6~hours, and a period of about 2.5~days. The shape of the transit in the discovery photometry was rather caused by the cadence of 30~minutes instead of a grazing geometry. In a time sampling that is high enough, the transit is U-shaped with a duration of 1.2~hours and a depth of about 1.8~ppt. It is a low-density planet of 3.5~Earth radii and about 11 Earth masses \citep{Marcy2014}. The planetary equilibrium temperature is 1095~K, as listed in the online catalog TEPCat \citep{Southworth2011}. A very precise ephemeris employing all available Kepler data is given by \cite{Holczer2016}.

\subsection{Kepler-63b}
Kepler-63b is the largest planet in the small planet sample targeted in this work. It is a giant planet intermediate in radius to Neptune and Saturn, and a mass lower than 0.4 Jupiter masses \citep{SanchisOjeda2013}. This planet produces the highest transit depth of our target sample with almost 4~ppt. The orbital period is comparably long with 9.4~days, causing an equilibrium temperature of about 900~Kelvin, somewhat cooler than the temperature of the majority of known close-in gas giants. The stellar spin axis and the planet orbital axis are mutually inclined. The planet orbits almost over the poles of its host star, revealing signatures of the repeated crossing of a long-lived polar spot \citep{SanchisOjeda2013}. The most current ephemeris employing all available Kepler data is drawn from \cite{Gajdos2019}.

\subsection{K2-100b}
K2-100b is a transiting Neptune-sized planet of $\sim$\,20 Earth masses and $\sim$\,1800~K equilibrium temperature. It is orbiting a member star of the open cluster Preasepe \citep{Mann2017}. It was observed by the Kepler satellite during the K2 mission in Campaign 5 in long-cadence mode and in Campaign 18 in short-cadence mode \citep{Mann2017,Barragan2019}. Independently, it was found in K2 data by \cite{Livingston2018}. The shallow transit depth of about 1~ppt was successfully detected from the ground with diffuser-assisted photometry \citep{Stefansson2018}. Its brightness of V=10.5~mag allowed for a radial velocity follow-up, which presented the first mass determination of a transiting planet in an open cluster \citep{Barragan2019}. Because of its current insolation, the authors also describe the planet to be likely evaporating, making it an interesting target to observationally study the photoevaporation of exoplanets.

\subsection{Multi-planet system K2-138}
The multiple planet system of K2-138 was discovered by \cite{Christiansen2018}. In photometric data of the Kepler satellite in its extended mission K2, the authors detected five sub-Neptune-sized planets close to a first-order resonant chain. The K-type host star is only moderately bright with V=12.2~mag. It shows a moderate level of stellar activity with a photometric variability amplitude of 1\,\% in the Kepler bandpass. The team of \cite{Lopez2019} succeeded in measuring the masses of the four inner planets by employing an extensive radial velocity monitoring of the host star, and found them to be of sub-Neptune mass. Because the planets are tightly packed, their mutual gravitational pull is expected to cause transit timing variations of about several minutes. However, for circular orbits, the amplitude of these variations should be below 10~minutes \citep{Lopez2019}. In contrast to the three objects mentioned above, the three planets b, c, and e of K2-138 do not have precise timing predictions. The most current ephemerides of \cite{Lopez2019} yield timing uncertainties in October 2019 from 41 to 133~minutes. We attempt to refine their timing prediction with our new observations.

\section{Observations and data reduction}
\label{sec_obs}

\begin{table*}
\caption{Overview of observations taken with the STELLA telescope. The columns provide the observing date, the number of the observed individual data points, the exposure time, the observing cadence, the dispersion of the data points as root mean square (rms) of the observations after subtracting a transit model and a detrending function, the $\beta$ factor (see Section \ref{sec_ana}), and the airmass range of the observations.}
\label{tab_overview}
\begin{center}
\begin{tabular}{llccccccc}
\hline
\hline
\noalign{\smallskip}
Object & Date & Filter & $N_{\mathrm{data}}$ &  $t_{\mathrm{exp}}$ (s) & Cadence (s) &   rms (mmag) &  $\beta$ &  Airmass \\
\hline
\noalign{\smallskip}
Kepler-94b & June 22, 2020 & r' & 188 & 70  & 96 & 1.15  &  1.00  & 1.06 - 1.38   \\
Kepler-94b & July 7, 2020  & r' & 128 & 70  & 97 & 1.45  &  1.00  & 1.09 - 1.97   \\
Kepler-94b & July 22, 2020 & r' & 136 & 70  & 97  & 1.32  &  1.43  & 1.10 - 1.96   \\
\hline
\noalign{\smallskip}
Kepler-63b & May 12, 2020 & g'  & 93  & 40  &  67  & 1.38  &  1.09  & 1.27 - 1.72 \\ 
Kepler-63b & Aug 5, 2020  & g'  & 157 & 40  &  67  & 2.40  &  1.59  & 1.09 - 1.53 \\ 
Kepler-63b & Oct 29, 2020 & g'  & 178 & 40  &  67  & 2.39 &   1.00  & 1.21 - 2.40 \\
\hline
\noalign{\smallskip}
K2-100b & Mar 9, 2020  & V & 347 & 15  & 41 & 1.77  &  1.07  &  1.01 - 1.24  \\
K2-100b & Mar 14, 2020 & V & 349 & 15  & 41 & 1.45  &  1.51  &  1.01 - 1.14  \\
K2-100b & Dec 11, 2020 & V & 270 & 15  & 41 & 1.98  &  1.03  &  1.01 - 1.37  \\
K2-100b & Jan 16, 2021 & V & 326 & 15  & 41 & 2.03  &  1.00  &  1.04 - 2.48  \\
K2-100b & Jan 21, 2021 & V & 310 & 15  & 41 & 1.49  &  1.04  &  1.03 - 2.39  \\
\hline
\noalign{\smallskip}
K2-138b & Oct 16, 2019 & r' & 135 & 60  & 90  & 1.10 & 1.00 &  1.29 - 1.55   \\
K2-138b & Oct 23, 2019 & r' & 117 & 60  & 90  & 1.25 & 1.11 &  1.29 - 2.16  \\
K2-138b\&e & Nov 18, 2019 & r' & 184 & 60  & 86  & 0.92 & 1.22 & 1.29 - 1.88    \\
K2-138b & Aug 12, 2020 & r' & 129 & 60  & 86  & 1.24 & 1.00 &  1.29 - 1.73 \\
K2-138b & Aug 19, 2020 & r' &  88 & 60  & 86  & 1.23 & 1.00 &  1.32 - 1.82 \\
K2-138c & Nov 27, 2019 & r' & 142 & 60  &  86  & 0.93 & 1.31 & 1.29 - 2.12   \\
K2-138c & Oct 23, 2020 & r' & 185 & 60  &  87  &  1.26 & 1.00  & 1.29 - 2.26   \\
K2-138c & Nov 17, 2020 & r' & 169 & 60  &  86  &  1.11 & 1.04  & 1.29 - 1.87  \\

\hline                                                                                                     
\end{tabular}
\end{center}
\end{table*}

The transit observations of this work have been obtained with the robotic 1.2m STELLA-I telescope and its wide field imager WiFSIP \citep{Strassmeier2004}. During all observations, we applied a mild defocus to achieve a width of the target point spread function of about 3~arcseconds. A detailed observing log is given in Table~\ref{tab_overview}. The data reduction of the imaging frames followed the procedure described in earlier transit photometry using the same instrumentation \citep{Mallonn2015,Mallonn2016}. The bias and flatfield correction was performed with the STELLA pipeline. Then we ran an aperture photometry using the code Source Extractor \citep{Bertin96}. The flux was extracted in apertures of different shape (circular and automatically adjusted elliptical, which is MAG\_APER and MAG\_AUTO in Source Extractor) and size, and we selected as the best version the version that minimized the point-to-point scatter in the final light curve. The same criterion of scatter minimization was used to define the best selection of comparison stars available in the field of view \citep{Mallonn2015,Mallonn2016}.

\section{Transit analysis}
\label{sec_ana}

Throughout this work, we employed the software tool JKTEBOP to model the data \citep{Southworth2005,Southworth2011}. The models consist of a planetary transit component and a component for light-curve detrending to account for photometric variations caused mainly by the atmosphere of Earth. The transit event is modeled through the transit parameters $a/R_s$, with $a$ as the semimajor axis of the planet orbit and $R_s$ as the stellar radius, the planet orbital inclination $i$, the ratio of planet-to-star radius  $k\,=\,R_p/R_s$, the transit mid-time $T_0$, and the two coefficients $u$ and $v$ of the quadratic limb-darkening law. Throughout this work, we do not fit for the limb-darkening coefficients, but fix them to theoretical values tabulated in \cite{Claret2012} and \cite{Claret2013}. All the planets invested here are modeled under the assumption of a circular orbit, which agrees with the corresponding discovery papers.

For detrending, we employ a low-order polynomial with time as its independent parameter. We also tested whether a linear combination of polynomials of other observational parameters such as the airmass, the width and elongation of the stellar point spread function, the CCD detector position, the peak count rate, or the count rate of the sky background can achieve a better fit to the variation in the light curves. However, in the vast majority of our data, a simple second-order polynomial over time minimizes the Bayesian information criterion (BIC), thus we consistently apply this detrending function to all light curves of this work. Previous transit observation with STELLA/WiFSIP were also found to be modeled best with this function \citep{Mallonn2015,Mallonn2016,Mackebrandt2017,Alexoudi2018}. For simplicity, we do not mention the three coefficients of the second-order polynomial as free-to-fit parameters in the description of all the fit exercises in the remainder of this work, but we emphasize here that every performed transit fit of this work contains these three coefficients as free-to-fit per individual light curve.

The aperture photometry software Source Extractor also provides an estimation of the uncertainty of the individual data points in the the light curves. It is based on the photon noise of the target and the sky background and the read-out noise. However, in reality, this uncertainty value often turns out to be slightly underestimated. For a more reliable uncertainty estimation, we make use of the point-to-point scatter in the light curves. We fit an initial transit plus detrending model to each light curve, reject outliers by a $4\,\sigma$ clipping, and adjust the individual photometric uncertainties of the data points by a common factor that results in a reduced $\chi^2$ of unity. We then calculate the so-called $\beta$ factor to inflate the photometric uncertainties for the second time, now taking into account the potential correlated noise in the data. See details on the $\beta$ factor in \cite{Gillon2006}, \cite{Winn2008}, and \cite{Mallonn2015}.

The uncertainties of the transit parameters are calculated with ``task 8'' and ``task 9'' of JKTEBOP \citep{Southworth2008,Southworth2011}. The former is a Monte Carlo simulation, which we run with 5000 iterations, the latter is a residual-permutation algorithm. We run both methods and employ the higher value as our final parameter uncertainty.

Ground-based observations typically show a smooth variability in the light curve that is caused by effects of the Earth atmosphere. As explained above, we model this trend in the data with a second-order polynomial over time. In the following, we model all data sets with two models, first a transit plus detrending model, and second a detrending-only model. For both, we calculate the BIC, and use their difference to decide whether either of the two models is significantly favored over the other. If $\Delta$\,BIC is smaller than -10, the transit model is favored, and for $\Delta$\,BIC of larger than 10, the transit model can be ruled out and the detrending-only model is significantly favored. For values between -10 and 10, the data cannot differentiate, that is, the photometric precision is insufficient to detect or rule out the transit signature. A summary of the derived transit depth $k^2$, the transit timing $T_0$, and the $\Delta$\,BIC value as indicator of the significance of the detection is provided in Table~\ref{tab_timings}.

\begin{table*}
\caption{Transit timings of the light curves of this work. Note the varying level of transit detection significance, indicated in the last column, $\Delta$\,BIC, and described in the text. The epoch is given regarding the following literature ephemeris references: Kepler-94b \cite{Holczer2016}; Kepler-63b \cite{Gajdos2019}; K2-100b \cite{Barragan2019}; K2-138c Equation~\ref{ephem_K2138c} this work.
}
\label{tab_timings}
\begin{center}
\begin{tabular}{llrccc}
\hline
\hline
\noalign{\smallskip}
Object & Date & Epoch & Mid time & $k^2$ & $\Delta$\,BIC \\
& & & [BJD$_{\mathrm{TDB}}$ +2 450 000] & [ppm] & \\
\hline
\noalign{\smallskip}
Kepler-94b & June 22, 2020 & 1622 & 9023.5280 $\pm$ 0.0020 & 1640 $\pm$ 303 & $-30.4$ \\
Kepler-94b & July 7, 2020  & 1630 & 9043.5861 $\pm$ 0.0082 & 1600 $\pm$ 700 & $-4.6$ \\
Kepler-94b & July 22, 2020 & 1634 & 9053.6326 $\pm$ 0.0027 & 1950 $\pm$ 582 & $-12.2$ \\
\hline
\noalign{\smallskip}
Kepler-63b & May 12, 2020 & 421 & 8982.6183 $\pm$ 0.0030 & & $-13.8$ \\ 
Kepler-63b & Aug 5, 2020 &  430 & 9067.5317 $\pm$ 0.0073 & & $-13.2$ \\
Kepler-63b & Oct 29, 2020 &  439 & 9152.4359 $\pm$ 0.0034 & & $-23.7$ \\
\hline
\noalign{\smallskip}
K2-100b    & Mar  9, 2020 & 1062 & 8918.4082 $\pm$ 0.0080 & 520  $\pm$ 304 & $-1.5$ \\
K2-100b    & Mar 14, 2020 & 1065 & 8923.4309 $\pm$ 0.0043 & 1150 $\pm$ 460 & $-6.6$\\
K2-100b    & Dec 11, 2020 & 1227 & 9194.6148 $\pm$ 0.0041 & 1206 $\pm$ 465 & $-6.1$\\
K2-100b    & Jan 16, 2021 & 1249 & 9231.4260 $\pm$ 0.0051 & 1007 $\pm$ 343 & $-8.2$\\
K2-100b    & Jan 21, 2021 & 1252 & 9236.4479 $\pm$ 0.0039 & 1014 $\pm$ 287 & $-12.5$\\
\hline
\noalign{\smallskip}
K2-138c    & Nov 27, 2019 & 302 &  8815.4105 $\pm$ 0.0032 & 1220 $\pm$ 430 & $-7.7$ \\
K2-138c    & Oct 23, 2020 & 395 &  9146.4943 $\pm$ 0.0057 & 655 $\pm$ 271 & $-5.4$ \\
K2-138c    & Nov 17, 2020 & 402 &  9171.4044 $\pm$ 0.0031 & 1065 $\pm$ 313 & $-9.5$ \\
\hline                                                                                                     
\end{tabular}
\end{center}
\end{table*}

\subsection{Kepler-94b}
\label{sec_Kep94}
\begin{figure}
 \centering
 \includegraphics[height=0.45\textwidth,angle=270]{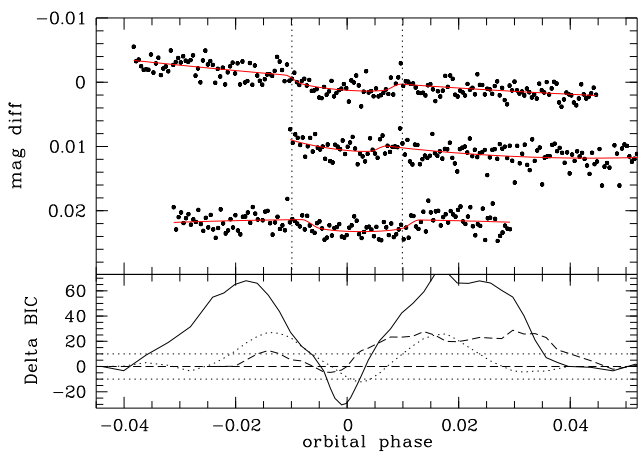}
 \caption{STELLA photometry of Kepler-94b. The orbital phase is calculated according to the ephemeris of \cite{Holczer2016}. The best-fit models with $k$ and $T_0$ as free parameter are overplotted as solid red lines. The predicted times of first and forth contact are marked with vertical dotted lines. Lower panel: Difference in the BIC of the best-fit model including the transit, in which the shape is fixed to the \cite{Marcy2014} transit parameters, and a model without the transit (detrending-only), when the transit is shifted along the x-axis. Values for light curves 1, 2, and 3 are given as solid, dashed, and dotted lines.}
 \label{plot_Kep94}
\end{figure}

\begin{figure}
 \centering
 \includegraphics[height=0.45\textwidth,width=0.31\textwidth,angle=270]{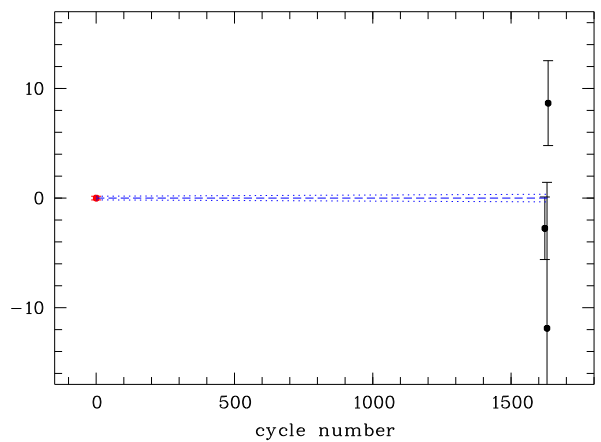}
 \caption{Observed minus calculated mid-transit times for Kepler-94b. Measurements of this work are plotted in black and the literature value of \cite{Holczer2016} in red. The ephemeris of \cite{Holczer2016} is given in blue, and the dashed lines present the 1\,$\sigma$ uncertainty range, which amounts to less than 1 minute at the transit epochs observed in this work. }
 \label{plot_OC_Kep94}
\end{figure}

We begin our transit light curve analysis with these targets of our sample for which the timing of the transits is precisely known. For Kepler-94b, the ephemeris determination by \cite{Holczer2016} is based on a long-time coverage of uninterrupted Kepler observations, thus the timing uncertainty at the epoch of our STELLA observations, epoch 1622, 1630, and 1634, amounts to only 0.3~minutes. We modeled our three data sets individually by keeping the fit parameters $a/R_s$ and $i$ fixed to the values of \cite{Marcy2014}, and fit for $k$, $T_0$, and the three coefficients of the detrending parabola. We detected the transit in the first and third data set, and obtained a tentative detection in the second light curve (Figure~\ref{plot_Kep94}). \cite{Marcy2014} estimated the transit depth to $k^2\,=\,1800$~ppm, which is significantly detected and confirmed in our first and third light curve with $1640 \pm 303$~ppm and $1950 \pm 582$~ppm, respectively. In the second light curve, we measured the depth with a lower precision to $k^2\,=\,1600 \pm 770$~ppm. In the estimated transit mid-time $T_0$ of the three individual light curves, we note some variation that is obvious by eye in the best-fit models in Figure~\ref{plot_Kep94}. However, comparing our measurements to the literature ephemeris in Figure~\ref{plot_OC_Kep94}, we found the variation to be not significant. The results for the first and second light curve agree to within 1\,$\sigma$, and the third light curve deviates by about 2\,$\sigma$. Due to our error bars of about 3, 12, and 4~minutes, we are  unable to refine the ephemeris.

The comparison of the BIC values between a transit plus detrending model, fixed to the \cite{Marcy2014} transit parameter, and a detrending-only model (Figure~\ref{plot_Kep94}) confirmed the significance of the transit signal detection in the first and third light curve by values of $\Delta$\,BIC\,$< -10$ at minimum, and the nonsignificant detection in the second light curve by a $\Delta$\,BIC of $-4.6$. We determined the $\Delta$\,BIC value not only for a transit model at the moment of predicted timing, but throughout our entire time series. To this end, we offset the mid-time $T_0$ of the transit model in discrete small steps, that is, we moved the transit model throughout the time series, and calculated $\Delta$\,BIC at each step (lower panel in Figure~\ref{plot_Kep94}). This exercise illustrates not only the capability of a light curve to detect the transit at a certain timing, but also to rule out the same transit signal at other timings covered by the observation. The more pronounced the minimum of the $\Delta$\,BIC at the moment of the detected transit signal, the stronger the variations of $\Delta$\,BIC over time and hence also its capability to rule out the signal at other moments in time. In the second light curve, we detect the transit signal only tentatively. This might be linked to the lack of observations before the predicted transit window, which makes it more difficult to clearly distinguish a transit dip from a minimum in the trend in the photometry caused by effects of the atmosphere of Earth. For the other two light curves, the power to clearly detect or rule out the transit drops quickly toward both ends. This drop of $\Delta$\,BIC toward zero starts at the timings at which the transit is still completely covered, but only a few data points are left before or after the transit to tighten the detrending parabola. The second light curve is also able to significantly rule out ($\Delta$\,BIC > 10) that the transit signal takes place at a later time at which it would be fully observed (Figure~\ref{plot_Kep94}).

A joint analysis of all three light curves yields an estimate of $k^2$ of $1490 \pm 230$~ppm. This deviates from the Kepler measurement of \cite{Marcy2014} by only 1.3\,$\sigma$ and constitutes a 6\,$\sigma$ detection of this shallow transit.

\subsection{Kepler-63b}
\begin{figure}
 \centering
 \includegraphics[height=0.45\textwidth,angle=270]{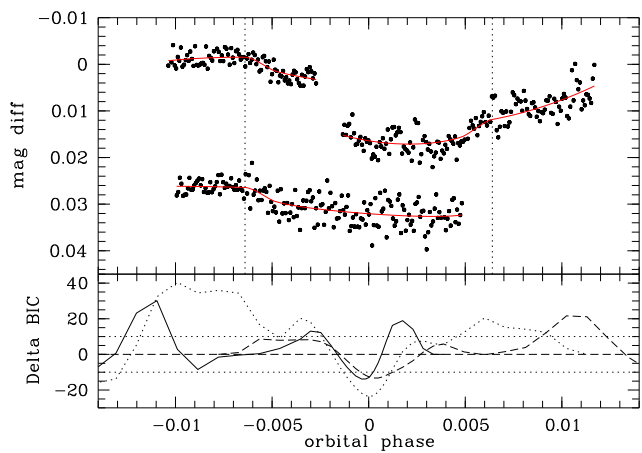}
 \caption{STELLA photometry of Kepler-63b. The orbital phase is calculated according to the ephemeris of \cite{Gajdos2019}. The best-fit models with $T_0$ as free parameter are overplotted as solid red lines. The predicted times of first and forth contact are marked with vertical dotted lines. Lower panel: Difference in the BIC of the best-fit model including the transit, in which the shape is fixed to the \cite{SanchisOjeda2013} transit parameters, and a model without the transit (detrending-only), when the transit is shifted along the x-axis. Values for light curves 1, 2, and 3 are given as solid, dashed, and dotted lines.} \label{plot_Kep63}
\end{figure}

We obtained photometric data for three transit events from Kepler-63b, but all observations covered the event only partially and lack either ingress or egress (Figure~\ref{plot_Kep63}). Our observations correspond to epochs 421, 430, and 439 with respect to the recent ephemeris of \cite{Gajdos2019}. The associated uncertainty of this ephemeris amounts to only 0.2~minutes.

We began the analysis by a modeling run of the three light curves individually. We fixed $a/R_s$ and $i$ to the values derived by \cite{SanchisOjeda2013} and fit for $k$ and $T_0$. We realized that in all three light curves, the depth $k^2$ is overestimated by 1 to 2\,$\sigma$ , with low precisions ranging from 1.2 to 2.9~ppt. A major cause for low precision and low accuracy is certainly that it is difficult to constrain the detrending function by partial transit measurements. In a joint fit of all three light curves with individual detrending per light curve but a common value for $k$, we determined $k^2$ to $5.99 \pm 0.87$~ppt, which is higher than the literature value of 3.9~ppt of \cite{SanchisOjeda2013} by 2.4\,$\sigma$.

We ran a second modeling version and now also kept  $k$ to literature values, that is the shape of the transit signal was totally fixed. We determined the timing to verify that we measured the transit signal at the precisely predicted time and compared transit plus detrending models of timings along the measurements with the BIC of the detrending-only model to estimate the significance of the detection of the transit signal in the partial transit photometry. Figure~\ref{plot_Kep63} provides the difference of the BIC values, which is below $-10$ at the predicted moment of the transit for all three transits, that is, in all three data sets the transit signal is significantly detected (Figure~\ref{plot_Kep63}).

\begin{figure}
 \centering
 \includegraphics[height=0.45\textwidth,angle=270]{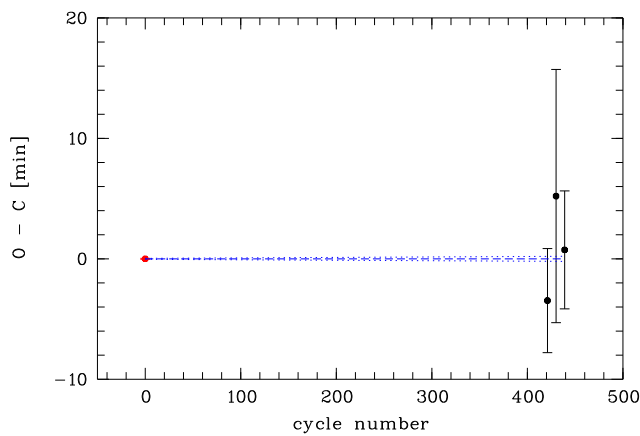} \caption{Observed minus calculated mid-transit times for Kepler-63b. The measurements of this work are plotted in black and the literature value of \cite{Gajdos2019} in red. The ephemeris of \cite{Gajdos2019} is shown in blue, and the dashed lines present the 1\,$\sigma$ uncertainty range, which amounts to less than 1 minute at the transit epochs observed in this work. }
 \label{plot_OC_Kep63}
\end{figure}

The timing accuracy of our three measurements is very satisfying. Figure~\ref{plot_OC_Kep63} shows that all measurements agree within 1\,$\sigma$ with themselves and with the prediction of \cite{Gajdos2019}. Similarly to our case of Kepler-94b, our timing precision from 4 to 10~minutes does not allow us to refine the ephemeris.

\subsection{K2-100b}
\label{sec_K2100}
\begin{figure}
 \centering
 \includegraphics[height=0.45\textwidth,width=0.35\textwidth,angle=270]{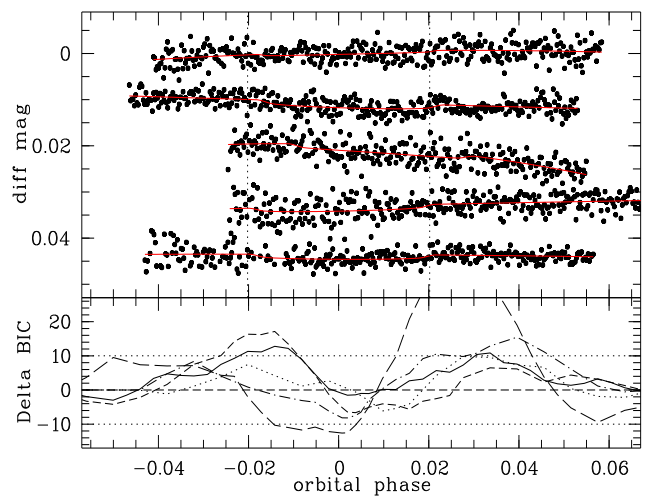}
 \caption{STELLA photometry of K2-100b. The orbital phase is calculated according to the ephemeris of \cite{Barragan2019}. The best-fit models with $k$ and $T_0$ as free parameter are overplotted as solid red lines. The predicted times of first and forth contact are marked with vertical dotted lines. Lower panel: Difference in the BIC of the best-fit model including the transit, in which the shape is fixed to the \cite{Barragan2019} transit parameters, and a model without the transit (detrending-only), when the transit is shifted along the x-axis. Values for light curves 1, 2, 3, 4, and 5 are given as solid, dashed, dotted, dashed-dotted, and long-dashed lines. The fifth light curve presents a significant detection crossing the $\Delta$\,BIC threshold, while the second to forth light curves present tentative detections with minimum $\Delta$\,BIC lower than $-5$.}
 \label{plot_K2100}
\end{figure}

We obtained five light curves of K2-100b at transit times predicted by \cite{Barragan2019}. According to this most recent literature ephemeris, our transit times correspond to epochs in the range from 1062 to 1252. In an individual transit fit per light curve with free-to-fit parameters $k$, and $T_0$, we achieve $k^2$ values from 520 to 1206~ppm with uncertainties from 287 to 465~ppm. All individual $k^2$ results agree to  within 1\,$\sigma$ with the literature value of 822~ppm \citep{Barragan2019}. One out of our five measurements reveals the transit signal individually by ruling out a zero-depth nondetection by more than $3\,\sigma$ ($k^2=1014\pm287$~ppm, Table~\ref{tab_timings}) and crosses the threshold of $\Delta\,\mathrm{BIC} < 10$ (Figure~\ref{plot_K2100}). Three other light curves yield tentative results with a transit depth deviating from zero by more than 2\,$\sigma$ and $\Delta$\,BIC smaller than $-5$. The very first of our observations shows the transit signature at lowest confidence, but this light curve also shows the minimum of $\Delta\,\mathrm{BIC}$ at the predicted transit time when we fixed the transit shape to the parameters of \cite{Barragan2019} and moved the transit signature through the observed time series. All best-fit transit depths and the $\Delta\,\mathrm{BIC}$ values are summarized in Table~\ref{tab_timings}. For completeness, we present the result of the timing measurements of all light curves in this table that were obtained in the fit that was used to derive the transit depth with both $T_0$ and $k$ as free fit parameters. However, we note that especially for the first light curve, a timing result also depends on the initial parameter value of the fit because the signal-to-noise ratio of the transit detection is very low. 

\begin{figure}
 \centering
 \includegraphics[height=0.45\textwidth,width=0.29\textwidth,angle=270]{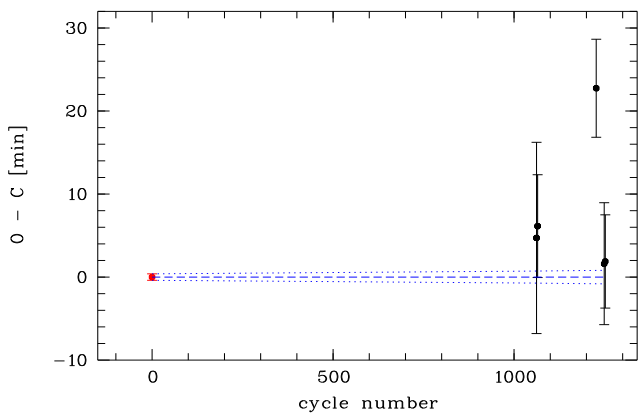}
 \caption{Observed minus calculated mid-transit times for K2-100b. The measurements of this work are shown in black and the literature value of \cite{Barragan2019} in red. The ephemeris of \cite{Barragan2019} is shown in blue, and the dashed lines present the 1\,$\sigma$ uncertainty range, which amounts to less than 1 minute at the transit epochs observed in this work. }
 \label{plot_OC_K2100}
\end{figure}

Figure~\ref{plot_OC_K2100} shows the timing deviations from the \cite{Barragan2019} ephemeris. While four out of five light curves agree with the prediction within  1\,$\sigma$, the third light curve estimates the transit signal to occur about 20 minutes later, which is a deviation of 3.8\,$\sigma$. We note here again that the transit signature is only tentatively detected in the individual light curve, therefore the individual timing measurements likewise need to be treated with caution. In a joint analysis of all five light curves, we achieve a depth $k^2\,=\,828\,\pm\,150$~ppm, which agrees very well with the value of \cite{Barragan2019} of 822~ppm and constitutes a detection of the shallow transit of K2-100b of 5.5\,$\sigma$.

\subsection{K2-138e}
\label{sec_ana_K2138e}
We analyzed the light curves of the multiplanet system K2-138 in the order of largest to smallest planet. Therefore we started with planet e. We obtained one observation of a transit of K2-138e at a time predicted by the ephemeris of \cite{Christiansen2018}. \cite{Lopez2019} published an updated ephemeris that concluded to a timing prediction different by only $\sim \, 5$~min. The photometric data are of high quality with a point-to-point scatter of 0.9~mmag in a cadence of 86~seconds and a mild $\beta$ factor of 1.2. A visual inspection of the light curve did not show any sign of the expected transit depth of 1.3~ppt and 3~hours duration. An attempt to fit for $T_0$ and $k$ while fixing $a/R_s$ and $i$ to the values of \cite{Lopez2019} ended with a JKTEBOP warning for not finding a good fit. Fixing $T_0$ also to the timing of \cite{Lopez2019} and only fitting for $k$ together with the detrending resulted in a null detection of a transit dip of $k = 0.0001 \pm 0.01$. That is, we can significantly rule out a transit signal of the shape of \cite{Lopez2019} at its predicted timing. 

\begin{figure}
 \centering
 \includegraphics[height=0.45\textwidth,angle=270]{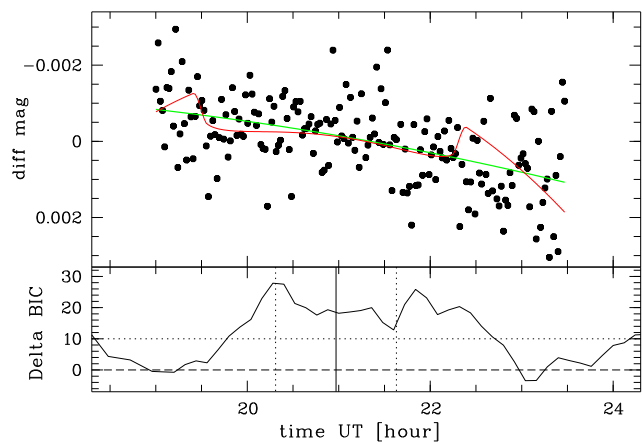}
 \caption{STELLA photometry of K2-138 at a predicted transit of planet e. The solid red line shows the best-fit model, including a transit model of fixed shape and time. The solid green line is a second-order time-dependent polynomial without a transit event. Lower panel: Difference in the BIC of the best-fit model including the transit, in which the shape is fixed to the \cite{Lopez2019} transit parameters, and a model without the transit (detrending-only), when the transit is shifted along the x-axis. At the predicted time of transit \citep[vertical solid line,][]{Lopez2019}, the transit model is significantly disfavored. Vertical dashed lines denote the $1\,\sigma$ uncertainty of the transit time prediction. }
 \label{plot_K2138e}
\end{figure}

To further search for the transit of planet e of given shape, we calculated the $\Delta\,\mathrm{BIC}$ value between the model combining the detrending function with the transit signal of fixed shape and the detrending-only model. We calculated the BIC difference moving the transit signal along our photometric time series. 
All transit timings that would have yielded a complete transit observation including ingress and egress give BIC differences larger than 10 (Figure~\ref{plot_K2138e}), which means that the model without a transit of the shape of \cite{Lopez2019} is significantly favored over a model including a transit event. This statement holds for a timing range of about $1.5\,\sigma$ before until about $2.5\,\sigma$ after the predicted transit timing of the most current ephemeris of \cite{Lopez2019}.

\subsection{K2-138c}
\begin{figure}
 \centering
 \includegraphics[height=0.45\textwidth,angle=270]{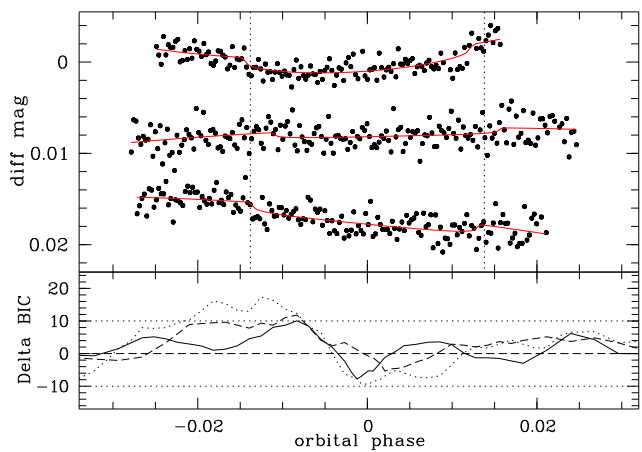}
 \caption{STELLA photometry of K2-138 at a predicted transit of planet c. The orbital phase is calculated according to the refined ephemeris of this work, Equation~\ref{ephem_K2138c}. The best-fit models with $k$ and $T_0$ as free parameter are overplotted as solid red lines. The predicted times of first and forth contact are marked with vertical dotted lines. Lower panel: Difference in the BIC of the best-fit model including the transit, in which the shape is fixed to the \cite{Lopez2019} transit parameters, and a model without the transit (detrending-only), when the transit is shifted along the x-axis. Values for light curves 1, 2, and 3 are given as solid, dashed, and dotted lines. For all three light curves, the minimum value of $\Delta$\,BIC suggests a tentative instead of significant detection.}
 \label{plot_K2138c}
\end{figure}

We obtained light curves of three transit events of K2-138c. They are shown in Figure~\ref{plot_K2138c}. In each individual data set, we tentatively detected a transit feature. When we searched explicitly for the transit shape of the most recent literature values \citep{Lopez2019}, the model including the transit was slightly preferred by the BIC over a detrending-only model by $\Delta$\,BIC = $-7.7$, $-5.4$, and $-9.5$, which means that none of the light curves reaches the threshold of -10 for a significant detection.
Modeling the light curves individually with JKTEBOP by fitting for $k$, $T_0$, and the detrending coefficients, we estimate the transit depth $k^2$ to be $1220\pm430$~ppm, $655\pm271$~ppm, and $1065 \pm 313$~ppm, respectively. All values agree within 1.5\,$\sigma$  with the literature value of $k^2\,=\,595$~ppm of \cite{Lopez2019}. 

\begin{figure}
 \centering
 \includegraphics[height=0.45\textwidth,angle=270]{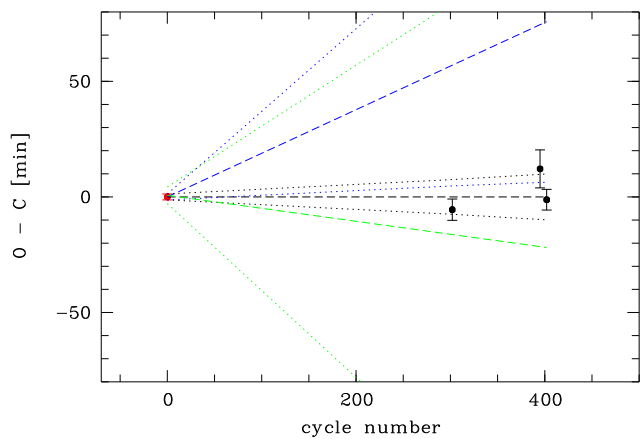}
 \caption{Observed minus calculated mid-transit times for K2-138c. The measurements of this work are plotted in black and the literature value of $T_0$ of \cite{Lopez2019} in red. The literature ephemerides of \cite{Christiansen2018} and \cite{Lopez2019} are presented in green and blue, respectively. The dashed line presents the timing prediction, and the dotted lines present the 1\,$\sigma$ uncertainty ranges. }
 \label{plot_OC_K2138c}
\end{figure}

On the individual transit mid-times, we achieve a precision of $\sim4.5$, 8, and 4.5~minutes. The timings of the first two light curves deviate from each other and with respect to a constant orbital period by $\sim$\,17~minutes, which corresponds to about 2\,$\sigma$ (Figure~\ref{plot_OC_K2138c}). This might either be caused by the low signal-to-noise ratio of the tentative transit detections, or might in principle also include transit timing variations within this compact multiplanet system, which could amount to 4~minutes \citep{Lopez2019}. All our individual timing measurements agree within 1\,$\sigma$  with the two published ephemerides of \cite{Christiansen2018} and \cite{Lopez2019}. Our precision is better than the timing uncertainties of their predictions, however, thus we can refine the ephemeris to 
\begin{equation}
T_c \,=\, 2457740.32182\,\pm\,0.00089\,+\,n\,\cdot\,3.559909\,\pm\,0.000017 ,
    \label{ephem_K2138c}
\end{equation}
given in BJD$_{\mathrm{TDB}}$, by fitting our data sets jointly, employing the orbital period $P$ as additional fit parameter, and including the zero-epoch timing measurement of \cite{Lopez2019}. The parameter $n$ denotes the number of orbital cycles that passed since the given transit zero epoch. 

The fit of the three data sets jointly yields a transit depth measurement of $k^2\,=\,752 \pm 192$~ppm, which denotes a $4\,\sigma$ detection of the transit feature and agrees to within $1\,\sigma$ with the transit depth of the two available literature references of \cite{Christiansen2018} and \cite{Lopez2019}.

\subsection{K2-138b}
\label{sec_ana_K2138b}
We obtained five observations of transit times predicted by \cite{Christiansen2018}. 
At the time of the scheduling of the observations in July 2019, this paper presented the most recent reference. Later in the same year, \cite{Lopez2019} published refined transit parameters and more precise ephemerides of the planets. Unfortunately, by that time it went unnoticed by us that the ephemeris of \cite{Lopez2019} predicted transit times earlier by 1.3 hours in 2019 and 1.7~hours in 2020, and we kept the scheduling to the ephemeris of \cite{Christiansen2018}.

\begin{figure}
 \centering
 \includegraphics[height=0.45\textwidth,width=0.35\textwidth,angle=270]{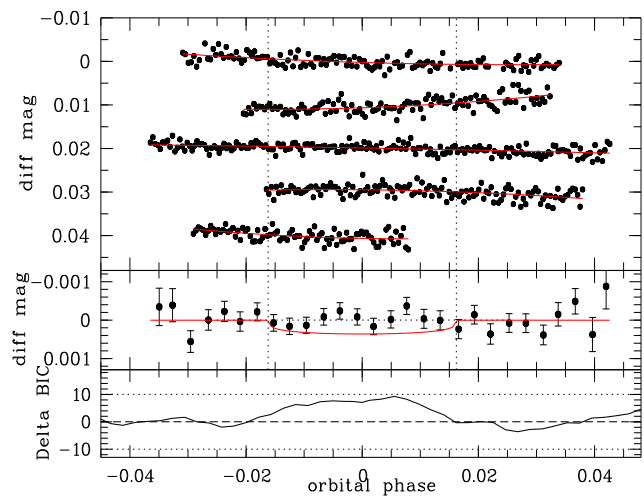}
 \caption{STELLA photometry of K2-138 at a predicted transit of planet b following the ephemeris of \cite{Christiansen2018}. Upper panel: Five STELLA photometric time series. Middle panel: Same photometry, detrended, phase-folded, and binned in 10 min intervals. The solid red line shows the \cite{Lopez2019} transit model. Lower panel: Difference in the BIC of the best-fit model including the transit, in which the shape is fixed to the \cite{Lopez2019} transit parameters, and a model without the transit (detrending-only), when the transit is shifted along the x-axis.  }
 \label{plot_K2138b}
\end{figure}

A transit fit of the combined data set of all five light curves with $k$, $T_0$, and $P$ as free parameters did not find a transit signature, but yielded a $k$ value very close to zero. The same was true when we fixed either $T_0$ or $P$.
We tested whether in principle we can rule out the presence of the transit feature at the time predicted by \cite{Christiansen2018} through the comparison of the BIC value of models with and without the transit signal. As shown in Figure~\ref{plot_K2138b}, the value of $\Delta$\,BIC is between 5 and 9 in a time window of about $\pm\,40$~minutes around the time preditiction of \cite{Christiansen2018}, which means that the presence of the transit is disfavored (in agreement with the prediction of the updated ephemeris by Lopez et al. 2019), but cannot be ruled out significantly. 

This data set of five observations is an interesting example to show that it would be misleading to invest the light curves \textit{\textup{after}} detrending. The middle panel of Figure~\ref{plot_K2138b} presents the phase-folded and binned light curve after the individual detrending of the data sets is subtracted in a prior step, overplotted with the transit signal of \cite{Lopez2019}. The data seem to be inconsistent with the transit shape, and seem to rule it out significantly. However, our comparison of the BIC values shows that the smooth trends cannot be distinguished from the transit, or vice versa.
We employ the data set of five observations for an injection-recovery test in the following Section~\ref{sec_dis}.

\section{Discussion}
\label{sec_dis}

We aim to test how shallow a transit signal can be to still be significantly detected in our light curves. To this end, we injected transit signals of various depths in certain light curves. For this exercise, we selected data sets of very good quality of our sample and a data set of lower photometric quality. We compared the results to another similar exercise using a light curve of simulated white noise. 

We began with one of the light curves of best photometric quality in our sample, the data obtained on November 18, 2019, for K2-138. The light curve has a point-to-point scatter of 0.92~mmag and a small $\beta$ factor of 1.22, that is, the mean of the photometric uncertainty is about 1.1~mmag after the error inflation described in Section~\ref{sec_ana}. Because our analysis favors the lack of a transit feature of planet e in this light curve (see Section~\ref{sec_ana_K2138e}), we used the original data instead of the light curve residuals after a model subtraction. We injected transit models of a given depth and chose the parameter values of $a/R_s$ and $i$ to yield a central transit with a duration of about 2~hours. The light curve has a duration of 4.5~hours, which means that the out-of-transit data points slightly exceed the in-transit measurements. We injected the transit at different timings along the data set to test whether the result varied, potentially caused by irregular mild features of correlated noise. After injection of a transit model, we modeled the resulting light curve with a transit plus detrending model, in which all the transit parameters were fixed to the input values, and with a detrending-only model. As in the previous sections, we considered the transit to be significantly detected if the transit plus detrending model was significantly favored by a $\Delta$\,BIC of -10. We find that this light curve can significantly reveal a transit of 1.3~ppt regardless of where it is injected as long as we guarantee for out-of-transit data toward both sides. For a strongly unequal number of pre- and post-transit data points (i.e., the transit is injected away from the center of the photometric time series), even shallower transits down to 0.7~ppt depth can be distinguished from a detrending-only model. 

For comparison, we created a simulated light curve with only white noise, featuring the same number of data points and a standard deviation of 1.12~mmag. For this simulated data set, the noncentral transit events can also be better distinguished from the detrending function than a transit event in the light curve center. The latter can apparently be better mimicked by a second-order polynomial over time than a transit signal with a longer out-of-transit observation at either side. Therefore we need a deeper transit to clearly differentiate its existence from the pure detrending function. The transit depth at which the threshold of -10 is reached is slightly shallower than for the real data, with 1.1 and 0.6~ppt for the central and the noncentral transits, respectively.

\begin{figure}
 \centering
 \includegraphics[height=0.45\textwidth,width=0.31\textwidth,angle=270]{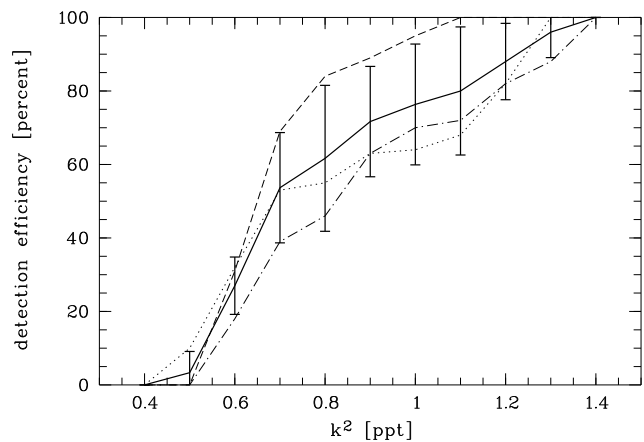}
 \caption{Transit signal detection efficiency for three selected individual data sets. We injected transit signals of various depths into the data and tested whether they were recovered significantly using the detection criterion of $\Delta\,\mathrm{BIC} < -10$. The detection efficiency for the individual light curves of November 18, 2019, June 22, 2020, and March 9, 2020, are shown with a dotted, dashed, and dash-dotted line, respectively. Their mean with standard deviation is shown as the thick solid line. Nearly 100\,\% detection efficiency is reached for signal depths of 1.3~ppt. }
 \label{plot_injrec}
\end{figure}

We repeated this injection-recovery analysis also for individual light curves of good quality of other stars, which is the light curve of Kepler-94 from June 22, 2020 (first light curve in Figure~\ref{plot_Kep94}) and of K2-100 from March 9, 2020 (first light curve in Figure~\ref{plot_K2100}). For these light curves, we first removed the best-fit transit model of Section~\ref{sec_Kep94} and Section~\ref{sec_K2100} and performed the test on the light curve residuals. Again, the depth which could be revealed by $\Delta\,\mathrm{BIC} < -10$ was dependent on where in the light curve it was injected, and it varied from 0.8 to 1.3~ppt. We can summarize that transit signals with a duration of 2~hours and a depth 1.3~ppt were revealed significantly by all three light curves, regardless of were in the light curve they were injected as long as they were fully covered by the data. The results for the detection efficiency, that is, the fraction of injected signals that could be significantly recovered, as a function of transit depth is shown in Figure~\ref{plot_injrec}.  

These injection and recovery tests agree with our findings in Section~\ref{sec_ana}. The transit of Kepler-94b with a depth of 1.8~ppt was detected significantly in the light curves that cover the event completely, including observations before and after the transit. The shallower transits of K2-100b and K2-138c with 0.8 and 0.6~ppt were not detected significantly in the majority of single observations, but yielded only tentative results that did not reach the threshold for significance. 

We also ran an injection-recovery test on the light curve with the lowest photometric quality in this work. This is the observation from August 5, 2020, of Kepler-63b. It shows a point-to-point scatter of 2.4~mmag and a substantial contribution of correlated noise, indicated by a $\beta$ factor of about 1.6. In a duration of about 3~hours, it contains 157 data points. While the transit of Kepler-63b was only covered partially, now we injected transit signatures that were fully covered. Thus, we designed a nongrazing, rather short transit with a duration of 1.4 hours for injection. Again, we placed it at different time positions in the light curve, all positions allowing for pre- and post-transit data points. Certainly caused by the substantial amount of correlated noise, the ability to recover the transit signal is very different for the different time positions of the injection in the light curve. It ranges from a transit depth of 1.4~ppt to 3.5~ppt.

Another step forward in this exercise is determining what a joint data set of multiple individual light curves is capable of detecting. To this end, an injection-recovery test was also run on the five light curves of K2-138b of this work. We injected transit light curves with durations of 2 hours with various depths ranging from 0.25 to 1.5~ppt. The capability of distinguishing significantly (delta BIC > 10) between transit plus detrending model from a detrending-only model varied over orbital phase. At all injected phases at which the transit event was fully covered by data points, a model of 1.0~ppt was recovered significantly. A 0.75~ppt model was detected in the majority of the covered phase space, a 0.5~ppt model crossed the threshold for a minor part of the injections. The detection sensitivity seems to roughly increase with the square root of the number of light curves, but the number of light curves available here does not allow for a detailed evaluation. The shallowest model of 0.25~ppt was not detected at all, which agrees with our finding of Section~\ref{sec_ana_K2138b} that only a tentative statement was possible for the presence of the transit signature of K2-138b with a depth of 256~ppm \citep{Lopez2019}.

This injection-recovery results goes in line with the final transit depth precision we reached for K2-138c and K2-100b of 192 and 150~ppm when three and five individual light curves, respectively, were analyzed jointly. In this way, transits with a depth even below 500~ppm appear to be detectable by repeated observations with ground-based 1-meter-sized telescopes. This generalized statement is also backed by other published measurements, for example, the 6\,$\sigma$ detection of the $\sim\,700$~ppm deep secondary eclipse of WASP-103b by \cite{Delrez2018}, for which more than ten individual observations of the Euler-Swiss telescope and the TRAPPIST telescope were employed. 

Our results for the transit detection capabilities of individual and multiple jointly analyzed light curves of the 1.2m STELLA telescope can probably be generalized to other professional observatories of this aperture size. Photometric time series of exoplanet transits observed with other telescopes show very similar properties, see, for example, data taken with EulerCam at the Euler-Swiss telescope \citep{Delrez2018,Lendl2017,Ciceri2016,Lendl2013}, data from KeplerCam on the Fred Lawrence Whipple Observatory 1.2 m \citep{Yee2020,Patra2017,Chan2011}, or data obtained with the 1m telescopes of the Las Cumbres Observatory \citep{Hartman2020,Dawson2019,Southworth2019,Dragomir2015}.

\section{Conclusion}
\label{sec_concl}

We described the capabilities of meter-sized telescopes, using the 1.2m STELLA telescope as representative case, of detecting shallow transits of (sub-) Neptune-sized exoplanets. The challenge of the detection is to distinguish the transit dip from a photometric trend that is inherent in ground-based differential photometry. We started the exercise with the observations of three exoplanets, Kepler-63b, Kepler-94b, and K2-100b, with a transit depth of 3.9, 1.8, and 0.8~ppt, respectively. For all these three targets, the shape and the timing of the transit was precisely given in the literature. While the transit depth of Kepler-63b seems large and comparably simple to detect, our available observations for this target posed the additional challenge that the transit event was only partly observed in all cases, making it more difficult to distinguish a transit feature from an underlying photometric trend. 

We treated an individual transit signal as significantly detected if the transit depth $k^2$ is different from zero by more than three times its uncertainty, and if the BIC difference of the best-fit transit-plus-detrending model and the detrending-only exceeds -10. Kepler-94b was detected significantly in two out of three individual observations, with depth and timing results being in $2\,\sigma$ agreement with the literature values or better. The third data set yielded only a tentative detection, likely caused by the lack of pre-transit data, which causes a more uncertain determination of the trend and therefore a lower ability to distinguish trend from transit feature.
The transit of K2-100b is only half in depth compared to Kepler-94b, and the transit feature could not be detected significantly in four out of five available light curves. However, these data sets individually provide tentative indications for the presence of the transit feature at the predicted timing. The fifth data set crosses our detection thresholds individually.  The majority of the fit results for depth and timing of the five light curves of K2-100b agree  within $1\,\sigma$ with the literature values. The joint analysis of all data sets revealed the targeted transit feature of $\sim 0.8$~ppt depth to more than $5\,\sigma$ confidence. The deeper transit feature of Kepler-63b could be revealed significantly individually in each of our three observations, despite the lack of a complete transit coverage. 

The other three targets of our target sample are planets for which the transit timing ephemeris already accumulated uncertainties from 41 to 133~minutes. We succeeded in refining the ephemeris of planet K2-138c. Each of the three observations of the 0.6~ppt deep transit of K2-138c yielded indications for its detection at timings that agreed within $2\,\sigma$ with each other. A joint fit of the three light curves yielded a significant $4\,\sigma$ detection of the transit signature, and the best-fit transit depth value agrees well with the literature values. 

We were able to rule out that the transit mid-point of K2-138e occurs in the time window of -1.5 to $2.5\,\sigma$ around the predicted transit time of \cite{Lopez2019}, pointing to a problematic ephemeris and a need for additional follow-up to regain the transit. The much shallower transit signal of K2-138b of 0.3~ppt transit depth could not be distinguished from a pure detrending model by our data, although we jointly used five high-quality light curves. 

Our injection-recovery tests have shown that individual light curves of high quality can significantly reveal transit signatures of 1.3~ppt depth when they are fully covered by the observations, including out-of-transit observations before and after the event. In some cases, our single observations could detect transits even down to 0.7~ppt depth. These results agree well with the significant detections of Kepler-94b, the significant nondetection of K2-138e at the predicted transit window, and the nonsignificant indications for the transits of K2-100b and K2-138c in individual light curves. For the data set with the  lower quality, the ability to significantly distinguish a transit from a smooth trend drops. Only transits as large as 3.5~ppt could be revealed.

\begin{acknowledgements}
This work made use of the SIMBAD data base and VizieR catalog access tool, operated at CDS, Strasbourg, France, and of the NASA Astrophysics Data System (ADS).
\end{acknowledgements}

%
\bibliographystyle{aa} 
\bibliography{shallow_bib.bib} 
%

 

\end{document}